%% file: ms.tex
\documentclass[prd,aps,showpacs,preprintnumbers,amsmath,amssymb,nofootinbib,showkeys,twocolumn,floatfix]{revtex4-1}

\pdfoutput=1

\usepackage{hyperref}
\usepackage{natbib}
\usepackage{subfiles}
\usepackage{aas_macros}
\usepackage{amsmath}
\usepackage{amssymb}

\usepackage{graphicx}
\usepackage{dcolumn}
\usepackage{bm}
\usepackage{cases}
\usepackage[lofdepth,lotdepth,caption=false]{subfig}

\newcommand{\snx}{\sigma_{nX}}
\newcommand{\vbar}{\bar{v}}
\newcommand{\vesc}{v_{esc}}
\newcommand{\Rstar}{R_{\star}}

\newcommand{\be}{\begin{equation}}
\newcommand{\ee}{\end{equation}}
\newcommand{\pn}{p_N(\tau)}
\newcommand{\Ncut}{N_{cut}}
\newcommand{\vn}{v_{N}}

\newcommand{\mess}{\frac{3(\vn^2-\vesc^2)}{2\vbar^2}}
\newcommand{\avg}[1]{\langle #1 \rangle}
\newcommand{\sqsixoverpi}{\left(\frac{6}{\pi}\right)^{1/2}}
\newcommand{\vevbsq}{\frac{\vesc^2}{\vbar^2}}
\newcommand{\nmax}{N_{max}}
\newcommand{\soneinf}{\sum_{N=1}^{\infty}}
\newcommand{\unit}[1]{\mathrm{#1}}

\newcommand{\Msun}{M_{\odot}}

\begin{document}
\title{Comment on ``Multiscatter stellar capture of dark matter''}

\author{Cosmin Ilie}
\email[E-mail at: ]{cilie@colgate.edu}
 \altaffiliation[Additional Affiliation: ]{Department of Theoretical Physics, National Institute for Physics and Nuclear Engineering,  Magurele, P.O.Box M.G. 6, Romania}
\author{Jacob Pilawa}%
\author{Saiyang Zhang}
\affiliation{ Department of Physics and Astronomy, Colgate University\\
13 Oak Dr., Hamilton, NY 13346, U.S.A.
}%

\date{\today}%

\begin{abstract}
    Bramante, Delgado, and Martin [Phys. Rev. D {\textbf{96}}, 063002(2017)., hereafter BDM17] extended the analytical formalism of dark matter (DM) capture in a very important way, which allows, in principle, the use of compact astrophysical objects, such as neutron stars (NS), as dark matter detectors. In this comment, we point out the existence of a region in the  dark matter neutron scattering cross section $(\snx)$ vs. dark matter mass $(m_X)$ where the constraining power of this method is lost. This corresponds to a maximal temperature ($T_{crit}$) the NS has to have, in order to serve as a dark matter detector. In addition, we point out several typos and errors in BDM17 that do not affect drastically their conclusions. Moreover, we provide analytical approximations for the total capture rates of dark matter particle of arbitrary mass for various limiting regimes. 
    
\end{abstract}

\maketitle

\section{Introduction}\label{sec:Intro}

Dark mater capture formalism was originally developed in the 1980s for the case of weakly interacting massive particles (WIMPS), the most common and well motivated dark matter (DM) candidate at the time. In~\cite{Press:1985}, Press and Spergel estimate the capture rates of WIMPS by the Sun, whereas Gould generalizes this to more arbitrary massive astrophysical objects, including the Earth in~\cite{Gould:1987resonant}. Both those seminal papers, which serve as the basis of most subsequent work on dark matter capture by astrophysical objects~(e.g.~\cite{Aaron:2015,Freese:2008cap,Ilidio:2014,Iocco:2008cap,Kouvaris:2010,Lopes:2014,Moskalenko:2007,Niu:2017,Taoso:2010}), are limited to the capture of DM that undergoes, on average, at most one collision with a nucleus while it traverses the star. In the early 2000's the opposite regime was analysed by Albuquerque et al.~\cite{Albuquerque:2001}. For a class of strongly interacting superheavy DM particles known as SIMPZILLAs, ~\cite{Albuquerque:2001} finds analytic expressions for the capture rates by the Sun and the Earth. The intermediary regime, when a DM particle would have to collide with nuclei inside a star an order unity number of times remained out of reach until BDM17~\cite{Bramante:2017}~\footnote{see also ~\cite{Dasgupta:2019juq,Bell:2020}} was published in 2017. The authors calculate closed form, analytic expressions for the capture rates of DM of arbitrary mass with arbitrary cross scattering section by arbitrary astrophysical objects, opening up numerous avenues of research. Therefore, the theory of dark matter capture is now an almost complete chapter. 

Below, we briefly summarize the multiscattering capture formalism of BDM17.  The total capture rate can be written as a series, summing all partial capture rates $C_N$ when the DM particle has been slowed down below the escape velocity by exactly $N$ collisions:
\be\label{eq:Ctot}
C_{tot}=\sum_{N=1}^{\infty}C_N.
\ee

BDM17 provide analytical formulae for $C_N$ under the simplifying assumption that the amount of energy lost by the dark matter particle in each scattering event is $\Delta E=\beta_+E_0$ can be approximated by replacing the general kinematic quantity $z$ with its average value over all collisions: $\langle z\rangle$. Here $\beta_+\equiv4m_X m/(m_X+m)$, with $m_X$ the DM mass and $m$ the mass of the target nucleus. 
Following BDM17 we find, for capture of DM by an object where the local DM density is $n_X$ and the dispersion velocity of the Maxwell-Boltzmann DM distribution is $\vbar$:
\begin{widetext}
\be\label{eq:CN}
C_{N}=\frac{1}{3}\pi R^{2} p_{N}(\tau) \frac{\sqrt{6} n_{X}}{\sqrt{\pi} \bar{v}}\left(\left(2 \bar{v}^{2}+3 v_{e s c}^{2}\right)-\left(2 \bar{v}^{2}+3 v_{N}^{2}\right) \exp \left(-\frac{3\left(v_{N}^{2}-v_{e s c}^{2}\right)}{2 \bar{v}^{2}}\right)\right),
\ee
\end{widetext}
where $v_N=\vesc(1-\avg{z}\beta_+)^{-N/2}$ and $p_N(\tau)$ being the probability to collide exactly with N nuclei per crossing. The optical depth is: $\tau=2\Rstar~\sigma~n_T$,where $\Rstar$ is the radius of the star, $\sigma$ is the cross section of the DM nucleus scattering, and $n_T$ is the average number density of nuclei inside the star. We find an additional overall factor of $1/3$, when comparing our result with Eq.~22 of BDM17.    
Two of the authors of this comment used the formalism of BDM17 to show that dark matter capture can impose an upper limit on the masses of the first stars in~\cite{Ilie:2019}, hereafter IZ19. While exploring the observable implications of IZ19 (see~\cite{Ilie:2020PopIII}), the authors of this comment identified another typo and several errors in BDM17. First, for the equation that gives the approximate capture rate after exactly $N$ scatters in the limit when $m_X\gg m$ and $v_{esc}\gg\vbar$ (Eq.~23 of BDM17), we find the relative sign of the parenthetical term $\frac{2A_N^2\vbar^2}{3\vesc^2}$ is actually $+$, not $-$ as in BDM17. Namely, Eq~23 of BDM17 should read:
\small
\be\label{eq:CNapprox}
C_{N} = \sqrt{24\pi}p_{N}(\tau)n_{X}GM_{\star}R_{\star}\frac{1}{\bar{v}} \left(1-\left(1+\frac{2 A_{N}^{2} \bar{v}^{2}}{3v_{esc}^{2}}\right) e^{-A_{N}^{2}}\right)
\ee
\normalsize

 In~\cite{Ilie:2019Erratum} we find, for the parameter space explored in IZ19, the term in Eq.~23 of BDM17 with the wrong sign is subdominant and thus has no effects. After pointing out our concerns specified above in a series of emails to the authors of BDM17, we explored further the results of BDM17. In the following sections, we report our findings. Our main finding is the identification of a region in the $(\snx)$ vs. $(m_X)$ parameter space where constraining DM nucleon scattering cross sections with neutron stars becomes impossible. This is because in that region the total capture rate becomes insensitive to both $\snx$ and $m_X$. Conversely, for a given ambient dark matter density, we find the maximum temperature a NS of a given mass has to have ($T_{crit}$) such that it can be used to constrain the neutron dark matter scattering cross section.  

\section{Results}
 
 \subsection{Analytical approximations of the total capture rates}\label{ssec:approx}
   Below we derive analytical approximations for the total capture rates, that will be used to validate our numerical results. The probability that a DM particle will collide exactly $N$ times as it crosses a star, $p_N$, was shown in IZ19 to have the following closed form: $\pn=\frac{2}{\tau^2}\left(N+1-\frac{\Gamma(N+2,\tau)}{N!}\right)$, where $\Gamma(a,b)$ is the incomplete gamma function. From the above form, we get the following approximations:
\begin{subnumcases}{\pn\approx}
\frac{2\tau^{N}}{N!(N+2)}+\mathcal{O}(\tau^{N+1}), & \text{if } $\tau\ll 1$ \label{eq:pnlt} \\
\frac{2}{\tau^{2}}(N+1)\Theta(\tau-N), & \text{if } $\tau\gg 1$\label{eq:pnht}.
\end{subnumcases}

In the next section, we discuss our reproductions of Fig.~2 of BDM17. Namely, we calculate the mass capture rate $m_XC_X$ as a function of $m_X$ numerically by adding the terms in Eq.~\ref{eq:Ctot}. Below we develop analytical closed form approximations for those sums.
We start with the general situation, where the coefficients $C_N$'s are given by Eq.~\ref{eq:CN}. Analyzing the limiting behaviour of the exponent:
\small
\begin{subnumcases}{R_v\equiv\mess\approx}
\frac{3}{2}(2^N-1)\frac{\vesc^2}{\vbar^2}, & \text{\!\!\!\!\!\!\!\!\!if } $m~\sim~m_X$ \label{eq:Rlowmx} \\
\frac{3}{2}N\frac{m}{m_X}\vevbsq, & \text{\!\!\!\!\!\!\!\!\!if } $m\ll m_X$\label{eq:Rhighmx}
\end{subnumcases}
\normalsize
we can expand Eq.~\ref{eq:CN}, finding for $R_v\gg1$, and $R_v\ll1$, respectivelly:
\small
\begin{subnumcases}{C_N\approx}
\frac{1}{3}\sqsixoverpi\pi R^2\pn n_X\frac{3\vesc^2+2\vbar^2}{\vbar},& \label{eq:CNbigR} \\
\frac{3}{2}\sqsixoverpi\pi R^2\pn\frac{n_X \vesc^4}{\vbar^3}\beta_+\avg{z}\left(N+N^2\beta_+\avg{z}\right)\label{eq:CNsmallR}&.
\end{subnumcases}
\normalsize
The conditions on $R_v$ can be recast as conditions on the ratio between the mass of the target nuclei~($m$), and the mass of the DM particle~($m_X$), in view of Eqns.~\ref{eq:Rlowmx} and~\ref{eq:Rhighmx}. 

For the total capture rates, or the sum in Eq.~\ref{eq:Ctot} we  obtain with analytical approximations in the appropriate limits for $C_{tot,N_{max}}\equiv\sum_{N=1}^{\nmax}C_N$, where the sum is cutoff at an arbitrary $N_{max}$. There will be generally two regimes, $\tau\ll 1$ (single scattering), and $\tau \gg 1$. For the latter we get, by using Eqns.~\ref{eq:pnht},~\ref{eq:CNbigR} and~\ref{eq:CNsmallR}:
\tiny
\begin{subnumcases}{C_{tot,\nmax}\approx}
\left(\frac{2}{3\pi}\right)^{1/2}\frac{\pi R^{2}}{\tau^{2}}n_{X}\frac{3\vesc^{2}+2\vbar^{2}}{\vbar}\nmax(\nmax+3), &\label{eq:CtotNmaxbigR} \\
\sqsixoverpi\frac{\pi R^{2}}{\tau^{2}}n_{X}\frac{\vesc^{4}}{\vbar^3}\beta_{+}\avg{z}\nmax(\nmax+1)\times&\nonumber
\\
\times (\nmax+2)\left(1+\frac{\beta_{+}\avg{z}}{4}(1+3\nmax)\right).&
\label{eq:CtotoNmaxsmallR}
\end{subnumcases}
\normalsize
The equations above hold only for $\nmax\leq\Ncut\sim\tau$. When $\nmax>\Ncut$, since the sums are converged, we simply replace $\nmax$ with $N_{cut}$.

In the regime where $m_X\gg m$ and $\vesc\gg\vbar$ we can calculate $C_{tot}$ analytically, both for multi-scatter ($\tau\gg1$) and single scatter ($\tau\ll1$) capture. For details see~\cite{Ilie:2020PopIII}. The main result we find is a closed form for the following sum: {\small{$
\sum_{N=1}^{\infty}\pn\left(1-\left(1+\frac{2 A_{N}^{2}\bar{v}^{2}}{3v_{esc}^{2}}\right)e^{-A_{N}^{2}}\right) =T_1-T_2-T_3$}}, where we define {\small{$T_1\equiv\soneinf\pn$, $T_2\equiv\soneinf\pn e^{-A_N^2}$}}, and {\small{$T_3\equiv\soneinf\pn\frac{2A_N^2\vbar^2}{3\vesc^2}e^{-A_N^2}$}}. For the single scatter case, one can immediately get $C_{tot}=C_{1}$ from Eqns.~\ref{eq:CNsmallR} or~\ref{eq:CNbigR}. In this case there is a transition between a DM mass independent capture rate (Eq.~\ref{eq:CNsmallR}) to one that depends on mass via $\beta_+$. For the multiscatter case, we start with the simplest of the three therms, $T_1$. In that case, since $\tau\gg1$, we have: $T_1=1-p_0(\tau)\approx 1$. For $T_2$ we find:
\be\label{eq:T2}`
\soneinf\pn e^{-A_N^2}=\frac{2e^{-k\tau}(-1+e^{k\tau}-k\tau)}{(k\tau)^2},
\ee
where we introduced the following notation: $k\equiv\frac{3\vesc^2}{\vbar^2}\frac{m}{m_X}$. For $T_3$ we obtain:
\be\label{eq:T3}
\soneinf\pn\frac{2A_N^2\vbar^2}{3\vesc^2}e^{-A_N^2}=\frac{4m}{m_X}\frac{e^{-k\tau}(-2+2e^{k\tau}-2k\tau-k^2\tau^2)}{k^3\tau^2}.
\ee

\subsection{Testing the Analytical approximations}\label{ssec:testignapprox}
\begin{figure}[!htbp]
\centering
\subfloat[Comparison of our results to BDM17 Fig.~2 left panel]{\label{fig:WDCapNvsA}
\includegraphics[width=0.95\linewidth]{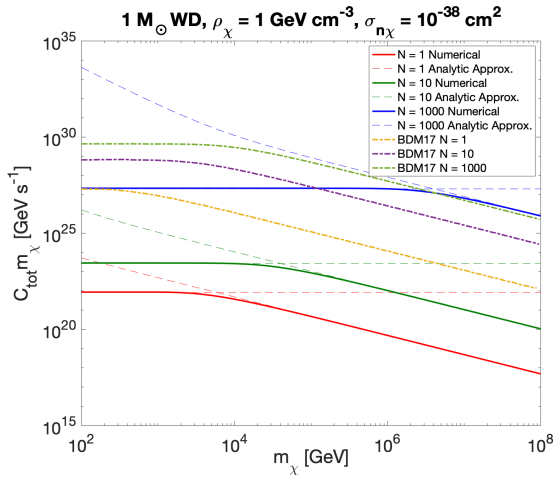}}\qquad
\subfloat[Comparison of our results to BDM17 Fig.~2 left panel after adjusting N. For details see discussion below.]{\label{fig:WDCapUsvsBDM}
\includegraphics[width=0.95\linewidth]{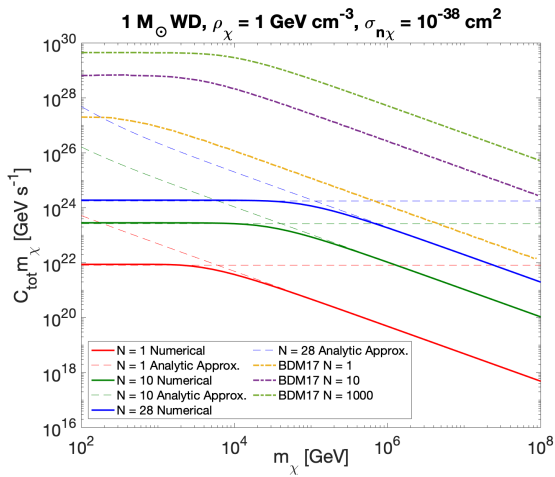}}\caption{Mass capture rate of dark matter on a constant density one solar mass white dwarf. For each case value of $\nmax$ is labeled in the legend. In adition to the values listed above each plot, we assumed $\vbar=220\unit{km}/\unit{s}$. Note that form factor effects lead to an effective DM scattering cross section with the target carbon nuclei enhanced by about four orders of magnitude, as explained in BDM17 (See Legend of their Fig.~2).}
\label{fig:WDCap}
\end{figure}

 We  validate our analytical approximations from Sec.~\ref{ssec:approx}, by applying them to the mass capture rates on a constant density white dwarf. As seen in Fig.~\ref{fig:WDCap} our analytical results presented in Eqns.~\ref{eq:CtotNmaxbigR} and~\ref{eq:CtotoNmaxsmallR} agree with the numerical results obtained by adding term by term $C_N$'s given by Eq.~\ref{eq:CN} up to an arbitrary $\nmax$.
We point out a severe discrepancy when comparing our results for the mass capture rates when a fixed arbitrary $\nmax$ is imposed, with those presented in BDM17 for the exact same set of parameters, as one can see from Fig.~\ref{fig:WDCap}. Using Eq.~\ref{eq:CtotoNmaxsmallR} we find the expected scaling with $\nmax$ of the mass capture rate in the regime where it is independent of $m_X$. Namely,  $C_{tot,\nmax} \sim \nmax(\nmax+3)$. We infer that the quoted value of $\nmax$ labeled as 1000 in Fig.~2 of BDM17 must be a typo, since the three values of the mass capture rate for $m_X=10~\unit{GeV}$ do not obey this scaling. We find the $\nmax$ the upper curve must be $28$ in order for the three curves to scale as $\nmax(\nmax+3)$. Even after correcting for this, we observe an overall discrepancy of about six orders of magnitude between our results an those of BDM17. We find a similar six orders of magnitude  discrepancy when we  reproduced the right panel of Fig.~2 of BDM17. Due to space constraints, we omit from including that here, but we are more than happy to make it available.   

\subsection{Probing heavy dark matter with old neutron stars}\label{ssec:NSDMdet}
\begin{figure}[!htbp]
\centering
\includegraphics[width=0.95\linewidth]{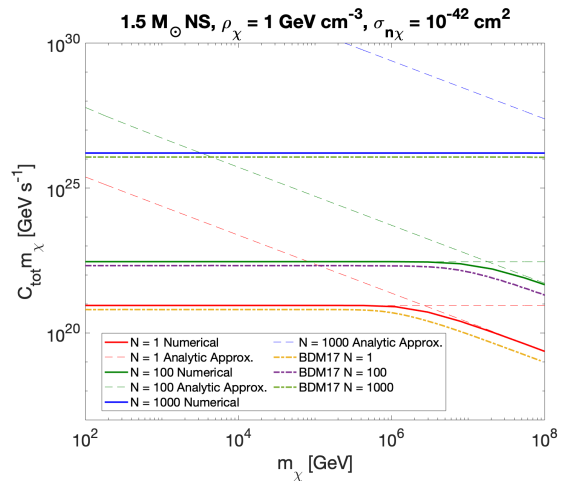}
\caption{Mass capture rate of dark matter on a neutron stars. Relevant parameters described above each figure. Comparison of our results to BDM17 Fig.~3 right panel.}
\label{fig:NSCap}
\end{figure}

We next move on to Figs.~3 and 4 of BMD17. Namely, we are interested in the effects of dark matter capture in neutron stars and follow the implementation of the general relativistic effects in the same way BDM17. The capture rate coefficients $C_N$ are enhanced to: $C_N\to\frac{C_N}{1-\frac{2GM}{Rc^2}}$,
and the escape velocity becomes: $\vesc\to\sqrt{2\chi}$, with $\chi=[1-(1-2GM/Rc^2)^{1/2}]$. Following BDM17, we focus first on the mass capture rate of DM on a neutron star, when adding term by term the GR enhanced $C_N$s up to an arbitrary cutoff $\nmax$. The results are presented in Fig.~\ref{fig:NSCap}. We again find that the mass capture rates from BDM17 are above our own estimates. In this case, however, the discrepancy is much milder.
Next, we explore the main topic of BDM17, the possibility of using the developed multiscatter formalism to constrain dark matter properties. Fig.~4 of BDM17 shows that upper-bounds on the neutron DM scattering cross section ($\sigma_{nX}$) as a function of the dark matter particle mass ($m_X$) coming from an old NS of a given temperature placed in a DM rich environment, such as the galactic center (GC), can be competitive, at low DM mass regime with direct detection experiments. Most importantly, for sufficienlty high DM mass, neutron stars can be used to probe below the ``neutrino floor,'' which is a major limitation of any direct detection experiment on Earth. 

\begin{figure}[!htbp]
\centering
\subfloat[Our full numerical (solid blue) and our analytical results (dashed light blue, and dashed brown lines)  vs. BDM17 (dashed green and blue lines).]{\label{fig:NSDmdetComp}
\includegraphics[width=0.95\linewidth]{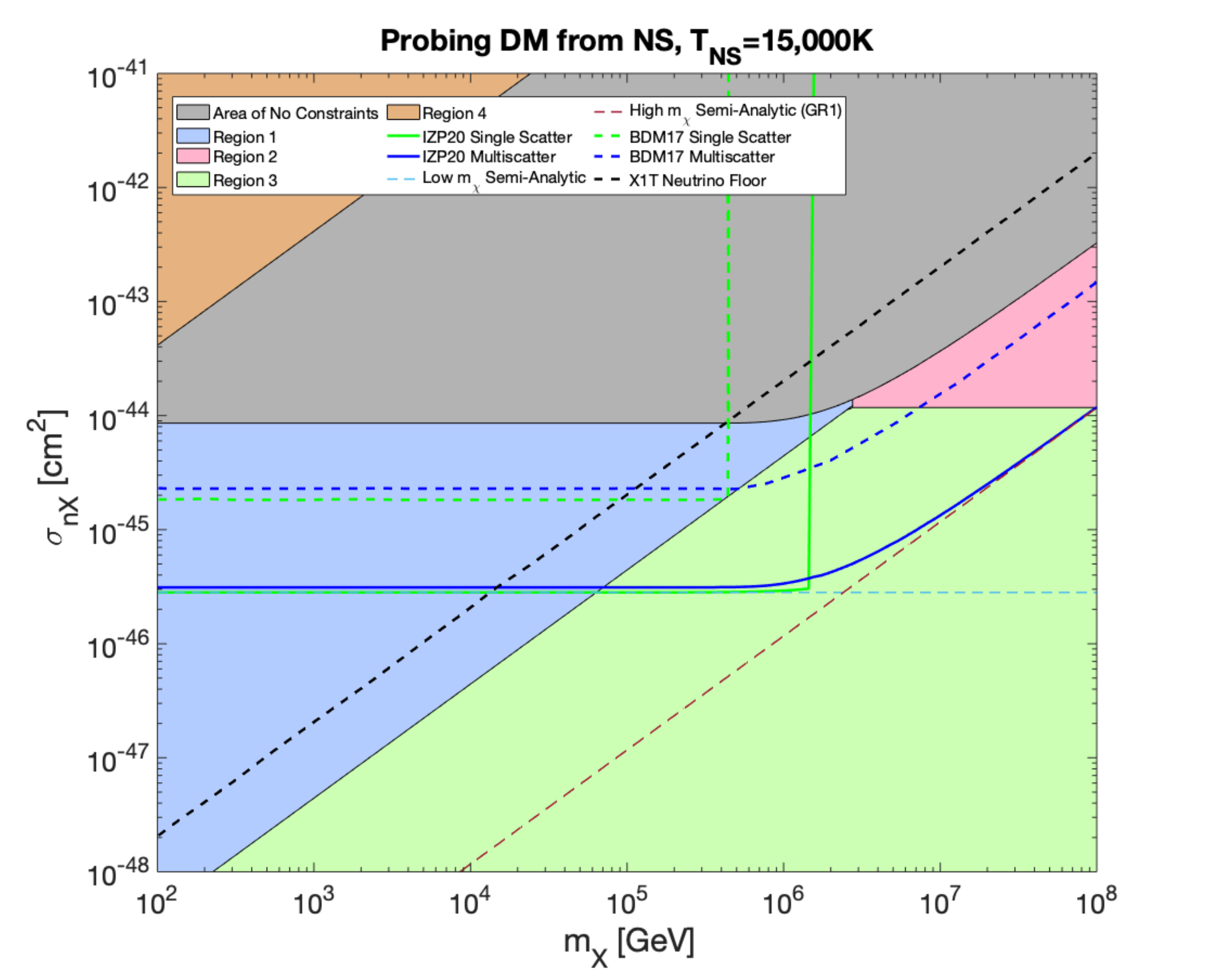}}\\
\subfloat[Bounds on neutron dark matter scattering cross section $\sigma_{nX}$ from the potential observation of a $\sim 23'000$~K (green line) neutron star in the GC region, $15000$~K (blue) and a 10000~K neutron star (purple).]{\label{fig:NSDmdetTwoTemps}
\includegraphics[width=0.95\linewidth]{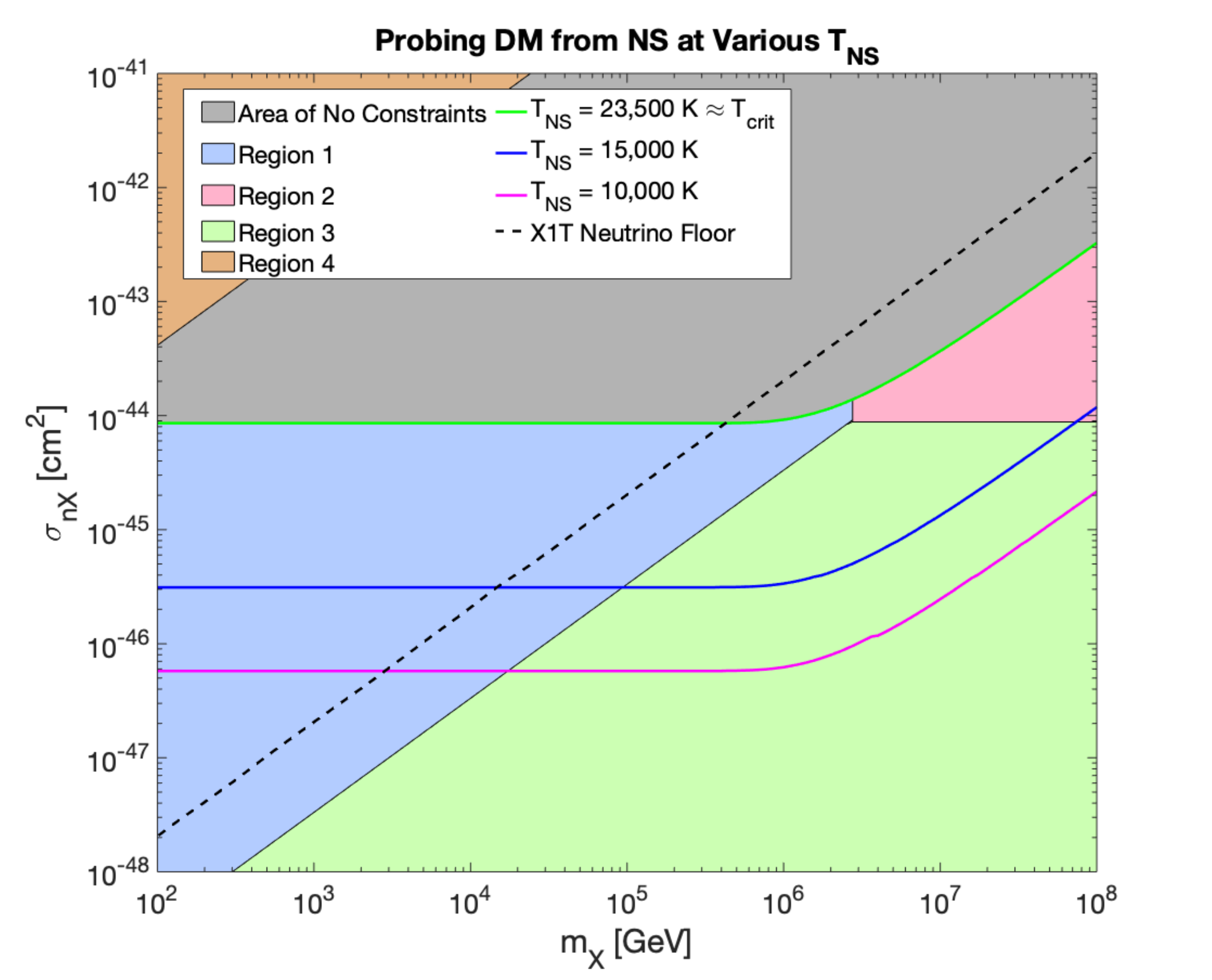}}
\caption{Constraints in neutron DM scattering cross section ($\sigma_{nX}$) and dark matter mass ($m_X$) parameter space coming from the potential  observation of a $1.5\Msun$ neutron star of various temperatures in the galactic center region. For the DM halo parameters we assume $\vbar=220$~km/s and a dark matter density at the center of the halo $\rho_X=10^3~\unit{GeV}\unit{cm}^{-3}$. For discussion about the various shaded regions labeled 1-4 and how we obtained our analytical approximation see discussion in the main body of the article.}
\label{fig:NSDmdet}
\end{figure}

The basis of the method proposed by BDM17 to use NS as dark matter detectors is briefly explained below. For old NS, an equilibrium between DM annihilation and capture can be reached on timescales $t_{eq}$ much less than the age of the neutron star itself. Moreover, typically, the captured DM thermalizes inside the NS on time scales smaller than $t_{eq}$. For details, see BDM17 Sec.~IV. Therefore, in order to calculate the luminosity generated inside the NS by captured DM annihilations, we only need to know the capture rates: $m_{X}C_{X}=L_{DM}$. Henceforth, we will denote by $C_{X}\equiv C_{tot}$, the total capture rate. Assuming the NS is in thermal equilibrium and that at $T_{NS}$ is maintained constant by DM heating, we have: 
\be\label{eq:NSThEq}
L_{DM}=L_{rad},
\ee
where $L_{rad}$ is the rate with which thermal energy is radiated at the surface of the neutron star:  $L_{rad}=4\pi R_{NS}^2\sigma_0T_{NS}^4$. Eq.~30 of BDM17 seems to just be a restatement of the equation above, with the only change the r.h.s is $L_{\infty}$, the apparent observed luminosity of the distant NS, not the thermally radiated power at the surface of the star. For completeness and clarity we reproduce Eq.~30 of BDM17: $m_XC_X=L_{DM}=4\pi\sigma_0R^2T_{NS}^4\left(1-2GM/Rc^2\right)^2$, where $\sigma_0$ is the Stefan-Boltzmann constant. The heating rate due to DM should be compared to the rate the NS is radiating energy at it's surface ($L_{rad}$), not to the rate observed by an infinitely distant observer ($L_{\infty}$), as implied by Eq.~30 of BDM17. This is our statement in Eq.~\ref{eq:NSThEq}. General relativistic effects such as time dilation and gravitational redshift lead to the following relationship: $L_{\infty}=L_{rad}(1-2GM/Rc^2)$~\cite{Thorne:1977ApJ,Hansel:2001}. Our point is that the bounds on DM parameters coming from the requirement of a NS of a given temperature $T_{NS}$ not overheating should actually be placed from our Eq.~\ref{eq:NSThEq}, not from Eq.~30 of BDM17, for the reasons explained above. Following this procedure, we can recast Eq.~\ref{eq:NSThEq} as:
\begin{widetext}
\be\label{eq:pnconstr}
\sum_{N=1}^{\infty}p_N(\tau)\left(1-\left(1+\frac{2A_N^2\vbar^2}{3\vesc^2}\right)e^{-A_N^2}\right)=4.6\frac{R_{NS}}{R_{10}}\frac{\vbar}{v_{220}}\frac{M_{1.5}}{M_{NS}}\left(1-\frac{r_S}{R_{NS}}\right)\frac{(T_{NS}/T_{30,000})^4}{\rho_X/\rho_{1000}},
\ee
\end{widetext}
where we introduced the following simplifying notations: $R_{10}\equiv10$~km, $v_{220}\equiv220$~km/s, $M_{1.5}\equiv1.5~M_{\odot}$, $r_S\equiv 2GM_{NS}/R_{NS}c^2$, $T_{30,000}\equiv 3\times 10^4$~K, and $\rho_{1000}=10^3~\unit{GeV}~\unit{cm}^{-3}$. Also note that we replaced the sign of the parenthetical term in the left hand side to $+$ form $-$, for reasons explained before (see discussion above Eq.~\ref{eq:CNapprox}). There are several things we note about Eq.~\ref{eq:pnconstr}. First, the left hand side is {\it{always}} less or equal to unity. This is obviously the case, since:
\scalebox{0.7}{%
$\sum_{N=1}^{\infty}p_N(\tau)\left(1-\left(1+\frac{2A_N^2\vbar^2}{3\vesc^2}\right)e^{-A_N^2}\right)\leq\sum_{N=1}^{\infty}p_N(\tau)\leq\sum_{N=0}^{\infty}p_N(\tau)=1$.}
This realization places a limit on the temperature of the NS, at a given dark matter density and halo dispersion velocity, for which the old neutron stars can be used as DM detectors. For NS with surface temperatures larger than this critical temperature, $T_{NS} \gtrsim T_{crit}$, one cannot use the NS as probes of dark matter, since the R.H.S. of Eq.~\ref{eq:pnconstr} becomes larger than 1. For $T_{crit}$ we get:
\small
\be\label{eq:Tcrit}
T_{crit}= 2.04 \times 10^{4} ~\unit{K} \left(1-\frac{r_S}{R_{NS}}\right)^{-1/4}\left(\frac{R_{10}}{R_{NS}}\frac{M_{NS}}{M_{1.5}}\frac{v_{220}}{\vbar}\frac{\rho_X}{\rho_{1000}}\right).
\ee
\normalsize
For the parameters used in BDM17, we get $T_{crit}\approx24,000$~K. Therefore, ignoring other heating~\footnote{such as kinetic heating~\cite{Bramante:2017}}, or cooling effects, and assuming the discovery of an old $1.5~\Msun$ NS at the GC that is not heating, only when its temperature is cooler than roughly $2.4\times 10^4~\unit{K}$ can it be used to place constraints on dark matter neutron scattering cross section. 
When $\left(1-\frac{r_S}{R_{NS}}\right)^{-1/4}$ is replaced by $\left(1-\frac{r_S}{R_{NS}}\right)^{-1/2}$, as inferred from Eq.~30 of BDM17, we get $T_{crit}\approx27,400$~K. The key points are: i) Eq.~30 of BDM17 is inadequate, and ii) both of those values of $T_{crit}$  are smaller than $30,000$~K, the value quoted in BDM17 for their exclusion limit presented in their Fig.~4. This is very puzzling since, as we demonstrated above, for $T_{NS} > T_{crit}$ one cannot use NS to place constraints on DM properties. In Fig.~\ref{fig:NSDmdetComp} we present the bounds we get for a NS with temperature $T_{NS}=1.5\times 10^{4}$~K and contrast those with the ones from Fig.~4 of BDM17. In Fig.~\ref{fig:NSDmdetTwoTemps} we show the bounds on the neutron DM scattering cross section implied by the observation at the GC of NS of temperatures ranging from $T_{crit}$ all the way down to $T_{NS}=10,000$~K.

We discuss below our main conclusions and results extracted from Fig.~\ref{fig:NSDmdet}, starting with the shaded regions. The gray shaded region corresponds to a region in the DM parameter space where old NS at the galactic center cannot be used as detectors. This region is defined as the constraints obtained by assuming a NS with $T_{NS}\approx T_{crit}$. For higher temperatures the r.h.s. of the equation that is used to place those constraints (Eq.~\ref{eq:pnconstr}) becomes larger than unity, whereas the l.h.s. is, by definition less than one. Next we move to the regions labeled 1-4. Region 1 (blue) corresponds roughly to $k\tau\gg1$ and $\tau\ll1$. Here, the single scatter approximation holds, and therefore the l.h.s. of Eq.~\ref{eq:pnconstr} can be approximated with the first term, for which $N=1$. Moreover, since $k\tau\gg1$, we can ignore the exponentially suppressed term, so the entire l.h.s. becomes $p_1(\tau)\approx2\tau/3$. Since $\tau\propto\sigma_{nX}$ and is independent of $m_X$, the constraints in this region correspond to constant values of $\sigma_{nX}$, as we can see from both our full numerical results and the analytical approximations. Region 2 (pink) corresponds to $\tau\gg 1$ and $k\tau\gg1$. Here we can use the analytical approximations of Eq.~\ref{eq:T2} and Eq.~\ref{eq:T3} to show that the l.h.s. of Eq.~\ref{eq:pnconstr} is approximately equal to $2k\tau/3\propto\sigma_{nX}/m_X$. For this reason the constraints approach $\sigma_{nX}\propto m_X$, as we transition to region 2. The green region (3) captures the transition between the two regimes described above. The brown shaded region (4) corresponds to the area where the term in Eq.23 of BDM17 that has the wrong sign (see discussion below our Eq.~\ref{eq:CNapprox}) would become important. In all other regions, that term is subdominant, and thus irrelevant. 

\section{Summary and Conclusion}
 In Sec.~\ref{sec:Intro}, we point out several errors and typos in BDM17, the paper that extended the DM capture formalism to the multiscattering regime. First, we identify a missing factor of $1/3$ in the equation giving the capture rate coefficient $C_N$ (Eq.~22 in BDM17 vs. Eq.~\ref{eq:CN} in this paper). Secondly, we note a wrong sign in the approximation of the capture rates coefficients $C_N$ when the escape velocity is much greater than the DM halo dispersion velocity, and the mass of the DM particle is much larger than the mass of the target nuclei (Eq.~23 of BDM17 vs. Eq.~\ref{eq:CNapprox} of this paper). In Sec.~\ref{ssec:approx}, we develop analytical approximation for the mass capture rates in various regimes (Eqns.~\ref{eq:Rhighmx},~\ref{eq:Rlowmx}, ~\ref{eq:T2}, and~\ref{eq:T3}). We proceeded to validate those analytical approximation in Sec.~\ref{ssec:testignapprox} and Sec.~\ref{ssec:NSDMdet}. In the process, we identify some errors in Figs.~2, 3, and 4 of BDM17. In Sec.~\ref{ssec:NSDMdet} we discuss the possibility of using neutron stars as DM detectors by requiring that the intrinsic luminosity due to captured DM annihilations does not exceed the rate with which the energy is radiated at the surface of the neutron star. In the process we identify a typo/error in Eq.~30 of BDM17. Independent of this subtlety, our most important finding is the identification of a critical temperature (see Eq.~\ref{eq:Tcrit}) above which a NS cannot act as a DM detector.

 \input{ms.bbl}

\end{document}

%% file: ms.bbl
%